\documentclass[conference, 10pt]{IEEEtran}
\usepackage{amsmath,amssymb,amsfonts}
\usepackage{algorithmicx}
\usepackage{algorithm2e}
\usepackage{graphicx}
\usepackage{textcomp}
\usepackage{xcolor}
\usepackage{listings}
\usepackage{bytefield}
\usepackage{multirow}
\usepackage{caption}
\usepackage{subcaption}
\usepackage{tikz}
\usepackage[normalem]{ulem}

\usetikzlibrary{graphs,arrows.meta,arrows,decorations.pathreplacing}

\RestyleAlgo{ruled}

\definecolor{codecomment}{rgb}{0,0.6,0}
\definecolor{codelinenumber}{rgb}{0.5,0.5,0.5}
\definecolor{codestring}{rgb}{0.58,0,0.82}
\definecolor{codeback}{rgb}{0.95,0.95,0.92}
\definecolor{codekeyword}{rgb}{0.0,0.0,0.95}

\begin{document}

\tikzset{
  treenode/.style = {align=center, inner sep=0pt, text centered,
    font=\sffamily},
  arn_n/.style = {treenode, circle, white, font=\sffamily\bfseries, draw=black,
    fill=black, text width=1.2em},
  arn_r/.style = {treenode, circle, red, draw=red, 
    text width=1.5em, very thick},
  arn_x/.style = {treenode, rectangle, draw=black,
    minimum width=0.5em, minimum height=0.5em},
  arn_s/.style = {treenode, circle, white, font=\sffamily\bfseries, draw=black,
    fill=black, text width=0.5em}
}

\title{Host-Based Allocators for Device Memory}

\author{\IEEEauthorblockN{1\textsuperscript{st} Oren Bell}
\IEEEauthorblockA{\textit{Washington University in St Louis} \\
St Louis, USA \\
oren.bell@wustl.edu}
\and
\IEEEauthorblockN{2\textsuperscript{nd} Ashwin Kumar}
\IEEEauthorblockA{\textit{Washington University in St Louis} \\
St Louis, USA \\
ashwinkumar@wustl.edu}
\and
\IEEEauthorblockN{3\textsuperscript{rd} Chris Gill}
\IEEEauthorblockA{\textit{Washington University in St Louis} \\
St Louis, USA \\
cdgill@wustl.edu}
}

\maketitle

\begin{abstract}
Memory allocation is a fairly mature field of computer science. However, we challenge a prevailing assumption in the literature over the last 50 years which, if reconsidered, necessitates a fundamental reevaluation of many classical memory management algorithms. We pose a model where the allocation algorithm runs on host memory but allocates device memory and so incur the following constraint: the allocator can't read the memory it is allocating.

This means we are unable to use boundary tags, which is a concept that has been ubiquitous in nearly every allocation algorithm. In this paper, we propose alternate algorithms to work around this constraint, and discuss in general the implications of this system model.
\end{abstract}


\section{Introduction}
\label{sec:intro}

This paper concerns itself with conventional dynamic memory management, i.e., algorithms to manage a heap of memory that clients can request and free blocks from at any time. The field of dynamic memory management could be said to have been started by Knuth \cite{knuth1968art}. Some 30 years of progress is well summarized by Wilson and Johnstone \cite{johnstone1998memory}\cite{wilson1995dynamic} in their comprehensive survey. Since then, further improvements have been made \cite{craciunas2008compacting}\cite{huang2010xmalloc}\cite{masmano2004tlsf}\cite{steinberger2012scatteralloc}, but overall the subject can be considered to be highly mature.


Most existing memory management algorithms include metadata as headers or footers within allocated blocks. In contrast, our system model assumes that the allocator is running on a host system, managing device memory. This may occur, for example, if part of a computation is offloaded to a GPU or FPGA device while the rest of the computation runs on a multicore processor. This in turn implies that the compute device cannot (conveniently and efficiently) access the memory it manages. Therefore, any data needed by the allocation algorithm cannot be stored in its allocated blocks.

We motivate this system model by considering the usecases surrounding how device memory is managed. In the current state of the art, memory allocation/deallocation is either done on the peripheral compute device in often proprietary device drivers. This causes memory allocators to be treated as a fixed black box.

However, different applications may have differing needs to manage their memory. The most optimal allocator may be high performant, real-time constrained, or have domain-specific characteristics catering to the application. Fixing the allocator choice in proprietary drivers and hardware denies developers the choice to optimize this aspect of their program through selective mapping of portions of the application and its supporting libraries to different computational devices.  Examples of such applications include drone cinematography~\cite{bonatti2020autonomous} and real-time hybrid simulation experiments in earthquake engineering~\cite{Condorietal2020}.

For our work, we assume that existing hardware and drivers are used to allocate arbitrarily large blocks of memory, but finer-grained memory allocation is then done in userspace by the host machine. As was mentioned previously, these allocators are constrained by the inability to read the memory they are managing and cannot store metadata in allocated blocks. We present alternative algorithms that overcome this constraint.

In Section~\ref{sec:free_lists}, we propose alternative measures to overcome this constraint of being unable to read managed memory. In Sections~\ref{sec:sequential_algorithms} and~\ref{sec:segregation_algorithms}, 
we present our updated alternative allocation algorithms. We compare the performance of one of our algorithms to the default CUDA memory allocation functions in Section~\ref{sec:evaluation}. Finally, we conclude in Section~\ref{sec:conclusion} with thoughts on implementations and usecases for this work.

\section{Background and Related Work}
\label{sec:existing}

Before introducing our new algorithms, we present a brief survey on existing memory allocation algorithms. Broadly speaking, these fall into 4 categories\cite{johnstone1998memory}.

\begin{description}
\item[Sequential Fits] \hfill \\ Free and allocated blocks of memory are linked together using header information within the allocated block itself. The resulting structure is called a free list. Examples include first fit, next fit, and best fit.
\item[Segregated Lists] \hfill \\ Portions of free data are grouped by size to aid in the lookup of a suitable block, including simple segregated storage, segregated fit, and TLSF.
\item[Buddy Systems] \hfill \\ Free blocks are only allowed to coalesce with a preassigned buddy. Generally sized in strictly powers of 2. Examples include binary buddy\cite{peterson1977buddy} and double buddy\cite{peterson1977buddy}\cite{wise1978double}
\item[SIMD Allocators] \hfill \\ Allocators designed for massively parallel programs \cite{huang2010xmalloc}\cite{steinberger2012scatteralloc}\cite{winter2021dynamic} where multiple threads may be requesting memory simultaneously. These are typically used in device code. In this paper, we consider memory is preallocated by host code, prior to a kernel launch. 
\end{description}

\subsection{Sequential Algorithms}
\label{sec:existingsequential}

Although there are a multitude of sequential algorithms, they all boil down to the same idea: all free blocks are kept in a free list, which is iterated through until a suitable block is identified for allocation. The only difference is what qualifies as a "suitable block".

The following are some of the most common sequential algorithms.

\begin{description}
\item[Best Fit] \hfill \\ The entire free list is iterated and the smallest block that is greater than the requested size is used. The search is terminated if a candidate block is exactly of the requested size.
\item[First Fit] \hfill \\ The first block that is large enough to accommodate the requested size is used.
\item[Next Fit] \hfill \\ Like first fit, except each traversal of the free list resumes from the last position, instead of starting over from the beginning.
\end{description}

In real world scenarios, tests have shown that next fit performs the best in terms of allocation time, although it suffers a fragmentation penalty compared to the others.

\subsection{Segregation Algorithms}
\label{sec:existingsegregation}

In segregation algorithms, multiple free lists are maintained for different size classes. There are three primary types.

\begin{description}
\item[Simple Segregated Storage] \hfill \\ Under this algorithm, every allocation request is rounded up to the next power of two. Free blocks are not split, and freed blocks are not coalesced. If a requested allocation has no available blocks in the free list for its size class, then a new block is created from the free heap.

\item[Segregated Fit] \hfill \\ This variation allows blocks to be split. Since not all blocks within a given size class are the same size, allocation requests are always serviced using the class size that is one degree of magnitude larger.

Since all blocks in the larger class size are guaranteed to be large enough to accommodate the request, it is unnecessary to perform iteration. These free lists are then treated like queues, only interacting with the front element. This prevents the frivolous search of a free list with no viable candidates, and reduces the allocation time to O(1). Additionally, it can perform deallocation in O(1) time, making it highly suitable for real-time systems, which require bounded execution.

One commonly used implementation is Doug Lea's Malloc\cite{lea1996memory}. It is a segregated fit algorithm that uses a combination of logarithm and linear spacing between its bins, achieving very low fragmentation in real-world evaluations\cite{johnstone1998memory}. It serves as the basis for the default allocator in the C programming language.

\item[Two-Level Segregated Fit (TLSF)\cite{masmano2008constant}\cite{masmano2003dynamic}\cite{masmano2004tlsf}] \hfill \\ Two-level segregated fit builds on the idea of segregated fit. As with segregated fit, TLSF has logarithmic size classes, but each size class is further divided into linear size classes. This retains the O(1) allocation and coalescence time complexity of segregated fit, but drastically reduces the fragmentation problem.
\end{description}

\subsection{Buddy Systems}
\label{sec:existingbuddy_systems}

Buddy Systems\cite{knowlton1965fast}\cite{peterson1977buddy} can be considered a subclass of segregation algorithms\cite{johnstone1998memory}, in that memory is sorted into size classes. However, they bear the additional constraint that freed blocks cannot be coalesced with any neighbor, only with a preordained buddy.

When an allocation request is made, it is rounded up to the size class. The heap is then split into two buddy blocks. The first is split further, and so on recursively until the desired size class is created. Like with segregated fit, each size class has a free-list allowing allocation in O(1) time.

After deallocation, if two buddies are both free, they are coalesced. This operation can be performed recursively up the binary heap, possibly taking O(logn) time.

This unbounded deallocation time means that applications with real-time constraints are often better served by a conventional segregated fit algorithm (such as TLSF).

Like with simple segregated storage, the commitment to block classes at strict sizes by this algorithm can result in a great deal of internal fragmentation within allocations. This can be somewhat assuaged by variants to the algorithm, listed below.

\begin{description}
\item[Binary Buddies\cite{knuth1968art}\cite{peterson1977buddy}] \hfill \\ The traditional and simplest implementation. It operates exactly as described above.

\item[Double Buddies\cite{peterson1977buddy}\cite{wise1978double}] \hfill \\  To partially resolve the internal fragmentation issue of conventional buddies, double buddies keeps two heaps with staggered class sizes, e.g., one heap with sizes of 2, 4, 8, ... and another with 3, 6, 12, ...

\item[Fibonacci Buddies\cite{hirschberg1973class}] \hfill \\ This approach assumes the heap size is a Fibonacci number. Since every Fibonacci number is the sum of two other Fibonacci numbers, blocks can be split recursively. This split is uneven, helping address the internal fragmentation caused by using strictly powers of 2. The ratio between consecutive size classes using Fibonacci buddies is approximately $\phi \approx 1.618$

\end{description}

\subsection{SIMD Allocators}
\label{sec:existingsimd}

This is a broad class of allocators designed for use specifically in SIMD systems, where many allocation requests may be made in parallel, creating risk for contention. A survey of currently available work is done in \cite{winter2021dynamic}.

Such allocators are often designed to avoid or eliminate locking overhead and are typically useful when threads in a GPU need to simultaneously allocate small amounts of memory in device code. This doesn't align with our model of host-based allocation code, and caters to a different use-case.

Our work is useful for parallel applications that are highly heterogeneous, intermixing both unaccelerated and accelerated tasks. The accelerated tasks operate on memory allocated prior to kernel launch. In such a system, all the memory allocation calls, both for host and device memory, would be done from host code.

Meanwhile, these SIMD allocators require execution of device code to run, leaving them outside the scope of the intersectional memory model we put forward. Furthermore, it has been demonstrated\cite{yang2018avoiding} that invoking device calls the manage CUDA memory causes implicit device-wide synchronization operations, impacting unrelated processes on the system. This presents a motivation to avoid device calls where possible. Managing device memory on the host achieves this.

\section{Alternatives to Free Lists}
\label{sec:free_lists}

Our paper explores a model in which the host device manages allocation of device memory. This prevents the allocator from reading the memory being managed, which presents new challenges not encountered in traditional memory allocation schemes. A key hurdle is that we cannot use boundary tags, a standard approach that was first mentioned by Knuth \cite{knuth1968art}.

\begin{figure}
    \centering
    \includegraphics[width=0.5\textwidth]{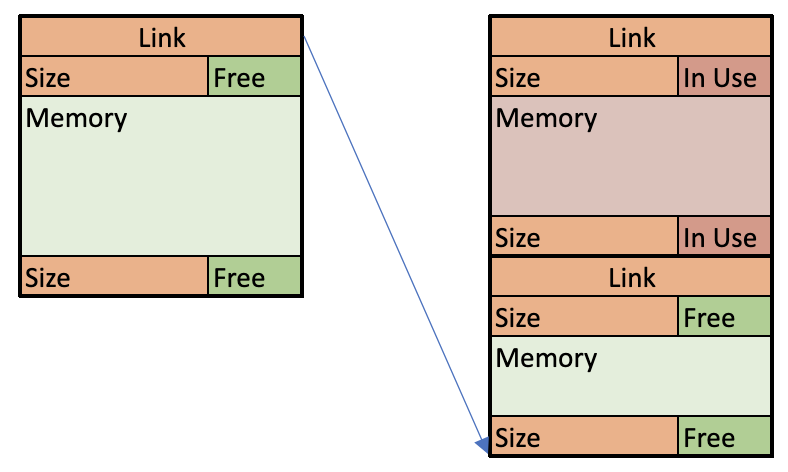}
    \caption{Demonstration of Boundary Tags}
    \label{fig:boundarytags}
\end{figure}

The traditional usage of boundary tags is illustrated in Figure \ref{fig:boundarytags}. A free list is formed by a linked list of blocks that are available for reuse. The header points to the next element in the free list, and the footer points to the header of the same block, which enables coalescence between two adjacent blocks in O(1) time. After a block is freed, one can subtract the footer size from the block's address to obtain the address of the header of the prior block. From there, one can check if that prior block is in the free list, and if so, the two will be coalesced. The next block in the free list can also be checked to see if it's adjacent in memory, possibly coalescing a total of three blocks.

These free lists are used in all Sequential Fit and Segregated Fit algorithms and a replacement mechanism will be needed to implement these in device memory.

\subsection{Arrays and Hybrid Array Lists}
\label{sec:free_listsarrays}

Array-based implementations of a free list for Sequential Fit algorithms using sub-pagesize allocations have high overhead, due to the potentially large number of small allocations. If we assume the smallest allocation size is 8 bytes, and an additional 8 bytes is needed to point at the free block, then the allocator would need a quantity of host memory equal to the device memory it is supposed to be managing. However, this is true even for conventional sequential fit allocations. Between the header and footer boundary tags, the worst case overhead is potentially 75\%.

With an array-based approach, we track all blocks, free or in-use. So each allocation of device memory only needs a single word (8 bytes) as overhead in host memory. We assume that we can spare a bit to indicate whether the block is free or not. Allocation and coalescence would naively be O(n) operations, corresponding to array insertion and deletion respectively, but we can reduce this by using Hybrid Array Lists (HALs)\cite{sara2020hybrid}

Our implementation is informed from prior literature\cite{sara2020hybrid}, and detailed in Appendix~\ref{app:hal}, which describes insertion, removal, and searching of a HAL. Two separate HALs are maintained: one to track free blocks, and another to track in-use blocks.

\begin{algorithm}
\caption{Memory Allocation}\label{alg:hal_alloc}
\KwData{$used\_list\_tail, free\_it, size \geq 0$}
$addr \gets free\_it.addr$\;
\uIf{$free\_it.size > size$}{
    \Comment{Split entry to only use space needed}
    $free\_it.address += size$\;
    $free\_it.size -= size$\;
}
\uElse{
    $remove(free\_it)$;
}
$it \gets search(used\_list\_tail, addr)$\;
$insert(it+1, \{addr, size\})$\;
\Return $addr$\;
\end{algorithm}

The allocation algorithm (Algorithm~\ref{alg:hal_alloc}) assumes that a candidate free block has already been identified: see Section~\ref{sec:sequential_algorithms} for discussion on that selection process. Given a candidate block in the free list, it must be moved to an in-use list, and possibly may be split if the block is larger than the request.

The free list is also guaranteed to always have at least one entry: the heap itself.

During deallocation (Algorithm~\ref{alg:hal_dealloc}), a block is removed from the in-use list and added to the free list. It will check the prior and subsequent blocks in the free list and perform coalescence if it is possible.

Partial deallocation is supported, e.g., freeing the last 2kB of a 10kB block. In the event of a partial deallocation, the in-use block will be shrunk, and otherwise it is simply removed from the in-use list.

When inserting into the free list, the previous and subsequent (aka left and right) blocks are checked and coalesced if either (or both) is adjacent to the block being inserted.

\begin{algorithm}
\caption{Memory Deallocation}\label{alg:hal_dealloc}
\KwData{$free\_list\_tail, used\_list\_tail, addr \geq 0$}

\Comment{Find chunk in used list}
$it \gets search(used\_list\_tail, addr)$\;
$size \gets it.size$\;

\Comment{Remove the chunk from the used list (or shrink it, in the case of partial deallocation)}\\
\uIf{$it.addr = addr$}{
    $remove(it)$\;
}
\uElse{
    $it.size = addr - it.addr$\;
    $size -= it.size$\;
}

$left \gets search(free\_list\_tail, addr)$\;
$right \gets left + 1$\;

\uIf{$left.address + left.size \geq addr$}{
    \uIf{$right.address \leq addr + size$}{
        $end \gets right.address + right.size$\;
    }
    \uElse{
        $end \gets addr + size$\;
    }
    $left.size \gets end - left.address$
}
\uElseIf{$right.address \leq addr + size$}{
    $right.size \gets right.address + right.size - addr$\;
    $right.address \gets addr$\;
}
\uElse{
    $insert(right,\{addr, size\})$\;
}

\end{algorithm}

\subsection{Bitmasks}
\label{sec:bitmasks}

Assuming O(n) searches are acceptable, one way to minimize overhead is through the use of bitmasks. In this approach, device memory is divided into its smallest allocatable size (e.g., 8 bytes), and the only overhead is a single bit in host memory to indicate whether that block is free or not. If we consider an allocation size lower bound of 8 bytes to be the worst case scenario, this creates an overhead of only ~1.5\% (one bit for 8 bytes). That produces 16MB of overhead for a gigabyte of device memory.

Allocation is done by finding a contiguous string of set bits in the bitmask corresponding to the desired size. For example, to allocate 1kB of memory, it is necessary to find a string of 128 bits that are all ones, incurring an O(n) cost for allocation, on the order of the address space size. The \texttt{clz} and \texttt{ffs} hardware instructions can be used to accelerate a bitsearch such as this.

All these bits must then be set to zero to indicate their corresponding blocks are in use. When the block is freed, they are all returned to ones (coalescence is implicit here). This means allocation and deallocation operations also incur an O(s) cost, on the order of the block size. 

Bitmasks have low overhead and relatively high time complexity on the order of allocation size.
Because of the low overhead, a potential use-case is to use bitmasks to manage a pool of small memory chunks, either of the same size (as in an object pool) or commingled varying sizes, as described above.

We can optimize the allocation search complexity by using the next-fit method. Prior work\cite{bays1977comparison}\cite{knuth1968art} has shown that next-fit improves on temporal performance over other sequential searches.

\subsection{Hash Tables}
\label{sec:hash_tables}

Another strategy is to use a hashtable, keyed by device memory address, to store all the information typically found in the header and footer of an allocated block.

The traditional algorithms for free-list traversal and coalescence operation would need to be modified to accommodate an extra step to lookup data in the hashtable. Additionally, the in-use blocks would also need to be tracked in the hashtable as well, as they also have necessary information in their header and footer.

With the use of hash tables, we aim to achieve O(1) allocation and deallocation, so co-mingling free and in-use blocks will not be a major performance concern.

The key for the hash table is the address of a block. The value contains i) the block size, ii) the address of the next block in a free list, iii) the address of a previous block in a free list, iv) the address of the previous block adjacent in address space. Thus we can create, in effect, free lists that span across a hash table. The ability to easily maintain multiple free lists is useful for segregated fit algorithms.

The primary downside of this approach is the overhead. Each entry contains 6 words (the key, block size, two references for a doubly linked list, one reference for a prior adjacent block, and a reference for a separate list linking collisions in the hash table). This means that in the worst case scenario, minimum allocation size of one word, the overhead is $\sim83\%$. This is comparable to the worst case overhead of 80\% seen in doubly linked free lists in conventional host-only algorithms. However, we only recommend this approach for applications with a fewer small allocations, such as those with a larger minimum block size, or any applications for which segregated fit algorithms are applicable.

\subsection{Comparison}

Below is a comparison of the three different strategies for replacing free lists. For allocation time of HALs and Hash Table lists, we ignore traversal when considering allocation. When these approaches are used in segregation algorithms (discussed later in Section~\ref{sec:segregation_algorithms}), they are often treated as stacks, with the head of the list always being a viable candidate. Bitmasks do not have this benefit. Searching for an allocation candidate will always require traversal of the memory space.

\begin{center}
\begin{tabular}{ | c | c | c | c | }
\hline
 Approach & WC Overhead & Allocation & Coalescence \\\hline
 HALs & 50\% & O(1) & O(n/m + m) \\
 Bitmasks & 1.5\% & O(n) & O(s) \\
 Hash Table & 87\% & O(1) & O(1) \\\hline
\end{tabular}
\end{center}

$n$ is the size of memory being managed. $m$ is the size of an array block in a hybrid-array-list. $s$ is the size of the allocated block.

Hybrid array lists are distinguished by their unbounded deallocation time. Even if an allocation algorithm doesn't require traversal to allocate a block, it may require traversal to coalesce. Since these hybrid array lists are intended to be sorted, entries are not guaranteed to remain in the same location in the list.

Allocation takes constant time, assuming all free blocks are eligible candidates, as is the case in segregation algorithms, which maintain multiple free lists for different size classes. Under this assumption, the last address in the list can be removed, which is as simple as decrementing a counter in the corresponding block. If the list must be traversed, as is the case of sequential algorithms (discussed in Section \ref{sec:sequential_algorithms}), then allocation becomes a O(n/m + m) process. This speed up in traversal, combined with lower overhead, makes HALs an ideal replacement for free lists where sequential algorithms are concerned.

Bitmasks always a require a traversal to identify an allocation candidate. This can be substantially sped up using the find-first-set command, which can search for 1-bit in a 64-bit word in a single machine-level instruction. This is still O(n), but offers a reasonable strategy for managing smaller object pools.

Storing linked lists in Hash Tables incurs a substantial overhead, so it is not recommended for smaller allocations. It does permit constant-time algorithms to retain their constant-time execution, which is crucial for real-time systems. This makes hash table stored lists the ideal replacement for free lists where segregation algorithms are concerned.

\section{Sequential Algorithms}\label{sec:sequential_algorithms}

All sequential algorithms are essentially the same, except for their stopping criteria. Below we include an algorithm for using best fit on device memory with hybrid array lists. This is followed by a description of the modifications needed for next-fit and first-fit.

\begin{algorithm}
\caption{Best Fit Allocation}\label{alg:best_fit}
\KwData{$free\_list\_tail, used\_list\_tail, size \geq 0$}
$it \gets free\_list\_tail$\;
$candidate \gets it$\;
$it++$\;
\While {$it.size \neq size \And it \neq free\_list\_tail$}{
    \If{$it.size < candidate.size \And it.size \geq size$}{
        $candidate \gets it$\;
    }
    $it$++\;
}

\Comment{Invoke Algorithm \ref{alg:hal_alloc} on candidate block}

\Return $allocate(used\_list\_tail, candidate, size)$\;

\end{algorithm}

Best-fit tries to find the smallest block that is large enough to accommodate the requested size. If a block with an identical size is found, the search can stop. However, it often requires traversing the entire list.

Contrarily, next-fit and first-fit select the first free block that is greater than or equal to the requested size. Next-fit is distinct from first-fit because it resumes the next allocation search where the last one left off. First-fit is memoryless and always restarts from the beginning of the free list.

All of the sequential algorithms use the same deallocation process, illustrated in Algorithm \ref{alg:hal_dealloc}.

\section{Segregation Algorithms}\label{sec:segregation_algorithms}

\subsection{Segregated Fit}

\begin{algorithm}
\caption{Allocate memory in segregated fit}\label{alg:segfit_alloc}
\KwData{$free\_lists, bitmask, hashtable, size \geq 0$}
\Comment{Find available free list large enough to accommodate}

$requested\_bins \gets 1 \ll ceil( log2( size ) )$\;
$order \gets ffs(requested\_bins)$\;
\Comment{Lookup head of candidate list in hashtable}

$it = ht[free\_lists[order]]$\;
$free\_lists[order] \gets it.next\_free$\;
$it.free = False$\;

\Comment{Split off surplus portion of block}

\uIf{$size < it.size$}{
$new\_block.size \gets it.size - size$\;
$new\_block.free \gets True$\;
$new\_block.addr \gets it.addr + size$\;
$new\_block.prev\_adj \gets it.addr$\;
$new\_block.prev \gets 0$\;
$it.size \gets size$\;

\Comment{Place new block at head of free list in its size class}
$bin\_idx \gets floor( log2( new\_block.size ) )$\;
$new\_block.next = free\_lists[bin\_idx]$\;
$new\_block.prev = 0$\;
$ht[new\_block.next].prev = new\_block.addr$\;
$free\_lists[bin\_idx] = new\_block.addr$\;
$bitmask \gets bitmask \  | \  (1 \ll bin_idx)$\;

$ht[ new\_block.addr + new\_block.size ].prev\_adj \gets new\_block.addr$\;

\Comment{Insert into the hashtable}
$ht[new\_block.addr] = new\_block$\;
}

$ht[it.addr] = it$\;

\Return $it.addr$\;
\end{algorithm}

In segregated fit, multiple free lists are maintained in size buckets for different powers of 2. Unlike sequential fit, the lists aren't searched. Instead, the first block is always selected. Given a requested size, it is rounded up to the nearest order of magnitude, and then the first block from that list is selected. This means (assuming no empty free lists) the selected block may be nearly 4x larger than the requested size, causing 75\% fragmentation, as discussed in prior literature~\cite{johnstone1998memory}~\cite{wilson1995dynamic}.

\begin{algorithm}
\caption{Deallocate memory in segregated fit}\label{alg:segfit_dealloc}
\KwData{$free\_lists, bitmask, hashtable, addr \geq 0$}
$it \gets ht[addr]$\;
$left \gets ht[it.prev\_adj]$\;
$right \gets ht[it.addr + it.size]$\;

\uIf{$left.free$}{
    \uIf{$right.free$}{
        \Comment{Coalesce all 3, erase it and right}
        
        $left.size += it.size + right.size$\;
        $ht[right.addr + right.size].prev\_adj \gets left.addr$\;

        \Comment{Update bitmask}
        
        \uIf{$right.prev = right.next$}{
            $unset\_bit \gets 2^{floor(log2(right.size))}$\;
            $bitmask \gets bitmask \And \lnot unset\_bit$\;
        }

        $remove(right)$\;
    }
    \uElse{
        \Comment{Coalesce left block, erase it}
        
        $left.size += it.size$\;
        $right.prev\_adj \gets left.addr$\;
    }
    
    \uIf{$it.prev\_free = it.next\_free$}{
        $unset\_bit \gets 2^{ceil(log2(it.size))}$\;
        $bitmask \gets bitmask \And \lnot unset\_bit$\;
    }
    $remove(it)$\;
    
    $it = left$\;
}
\uElseIf{$right.free$}{
    \Comment{Coalesce right block}
    
    $it.size += right.size$\;
    $ht[right.addr + right.size].prev\_adj \gets it.addr$\;
        
    \Comment{Update bitmask}
    
    \uIf{$right.prev = right.next$}{
        $unset\_bit \gets 2^{floor(log2(right.size))}$\;
        $bitmask \gets bitmask \And \lnot unset\_bit$\;
    }
    
    $remove(right)$\;
}

\Comment{Put coalesced block in free list in its size class}

$bin\_idx \gets floor( log2( it.size ) )$\;
$it.next = free\_lists[bin\_idx]$\;
$ht[it.next].prev = it.addr$\;
$it.prev = 0$\;
$free\_lists[bin\_idx] = it.addr$\;
$it.free = True$\;
$ht[it.addr] = it$\;
$bitmask \gets bitmask \ |\  2^{bin\_idx}$\;

\end{algorithm}

All lists are initialized pointing at a null block: an invalid block in the hashtable is meant to indicate the end of a free list. Accompanying this array of lists is an availability bitmap that can indicate which lists are non-empty. If the free list for a desired size class is empty, the next eligible list can be found by using the find-first-set (ffs) bit operation on the bitmap.

Our implementation stores the free lists in a hashtable, as discussed in Section \ref{sec:hash_tables}. Algorithm \ref{alg:segfit_alloc} and \ref{alg:segfit_dealloc} show pseudocode for allocation and deallocation, respectively.

\subsection{Two-Level Segregated Fit for Device Memory}

Two-Level Segregated Fit\cite{masmano2008constant}\cite{masmano2003dynamic}\cite{masmano2004tlsf} (TLSF) is a real-time dynamic memory allocation algorithm. Its primary benefits are reduced fragmentation and O(1) allocation and deallocation. It is an extension of segregated fit, dividing class sizes not only logarithmically, but also linearly, in a two-tiered system.

\begin{figure}[!h]
    \centering
    \includegraphics[width=\linewidth]{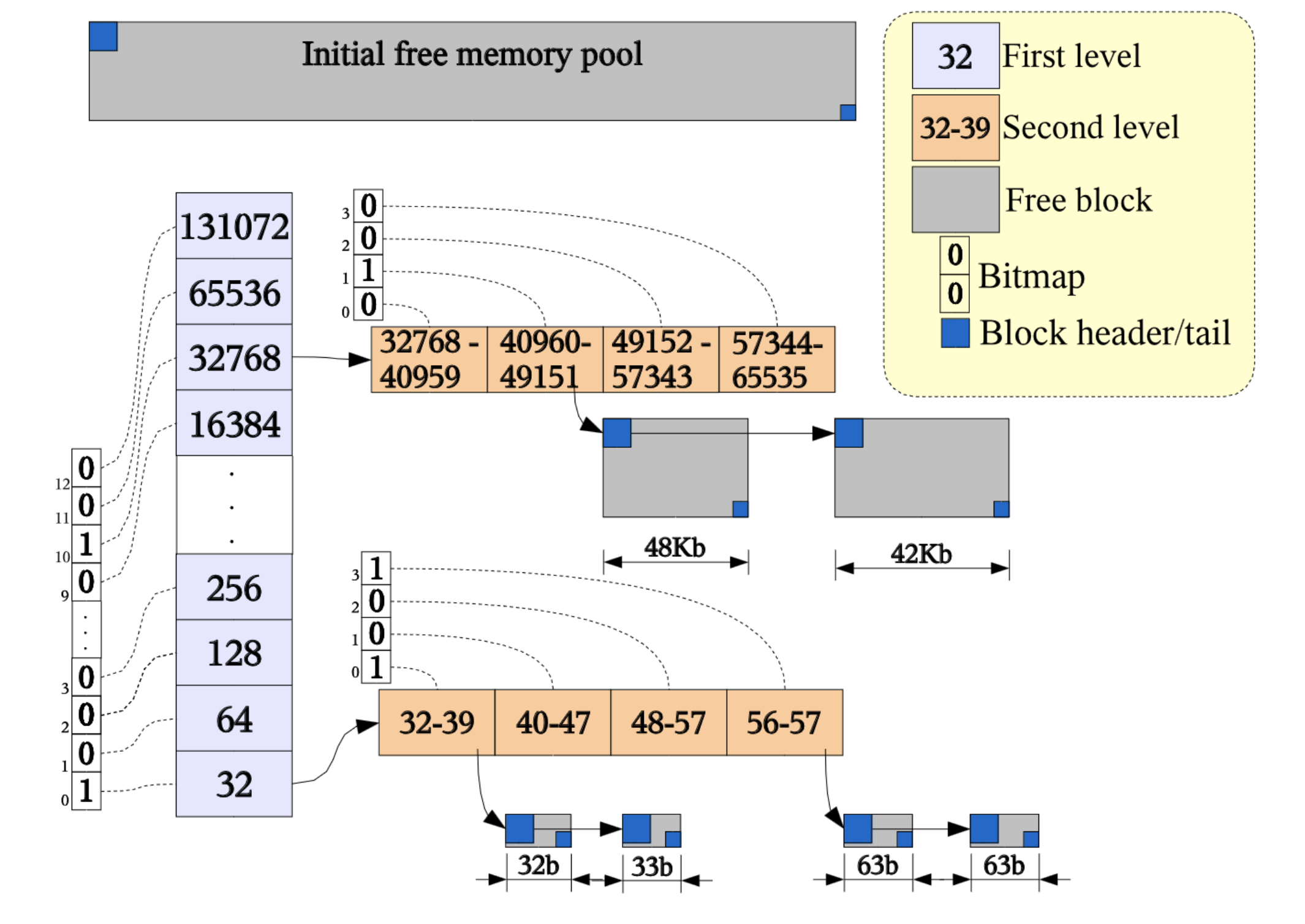}
    \caption{TLSF Data Structure\cite{masmano2003dynamic}}
    \label{fig:tlsf}
\end{figure}

The allocation and deallocation steps are functionally the same, except the lookup step requires consulting 2 bitmasks. One is logarithmic and functions just as detailed in Algorithm \ref{alg:segfit_alloc} and Algorithm \ref{alg:segfit_dealloc}. TLSF differs in that this points at another bitmask which subdivides the space linearly, as illustrated in Figure \ref{fig:tlsf}. This tier then points to a free list of blocks in that size class. If the free list is non-empty the linear bitmask will have a corresponding 1 set. If the linear bitmask is non-zero, then the logarithmic bitmask will have a corresponding 1 set.

Free list manipulations are done the same way as in segregated fit, so Algorithms~\ref{alg:segfit_alloc} and~\ref{alg:segfit_dealloc} only need to be modified to account for this two-tiered lookup process and bitmask manipulation.

\subsection{Hybrid Allocators and Object Buffers}
Hybrid approaches may also be employed, using object buffers to manage object pools for allocations smaller than a page ($<$4kB), and segregated lists for larger allocations.

Object pools allow for the efficient allocation of vast amounts of small allocations, but are not a practical approach for larger allocations. A hybrid approach can permit the best of both worlds.

But it is important to consider the practical use of such a hybrid approach in this context. A host-based allocator for device memory is unlikely to allocate large numbers of small objects, but rather large memory blocks that are passed to a kernel call for a hardware-accelerated device. If small allocations are created, they would be intermediary memory allocated by device code, likely using a SIMD allocator such as XMalloc\cite{huang2010xmalloc} or ScatterAlloc\cite{steinberger2012scatteralloc}. These allocators are designed to handle many allocation requests in parallel without the need for locks, a usecase that doesn't apply to the host-based model we have presented in our paper.

So, our usecase would be best served by limiting the smallest granularity to a medium sized value, such as 4kB, and exclusively using a segregated fit algorithm. 

\section{Evaluation}\label{sec:evaluation}

Evaluating the nuanced use cases that this work targets is challenging.
SIMD allocators do not offer a proper target for comparison since our allocators are run from the host and not intended for parallelization. Many good memory allocator benchmarks also exist for host memory\cite{leijen2019mimalloc}, including real-world applications. However, they too are unsuitable for comparison since the device memory we are allocating is inaccessible.

Instead, we compare our work to the default cuda allocator. We adapted the malloc-large benchmark from the MiMalloc benchmark suite\cite{leijen2019mimalloc} to use an implementation of our segregated fit algorithm. This algorithm randomly allocates and frees blocks 1kB and 16MB in size, several thousand times.

\begin{figure}
    \centering
    \includegraphics[width=\linewidth]{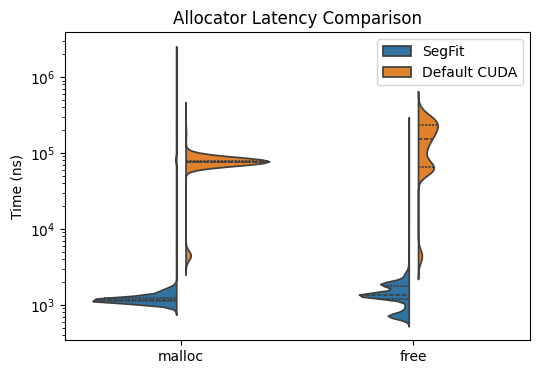}
    \caption{Latency of malloc and free}
    \label{fig:latencies}
\end{figure}

As a baseline, we also ran the benchmark with ordinary calls to cudaMalloc and cudaFree. The order and size of the alloc and free operations was consistent between our work and the baseline. Latency comparisons between the two are shown in Figure~\ref{fig:latencies}.

Our work outperforms the default CUDA allocator by 2 orders of magnitude. There are some outliers during program initialization where the segeregated fit malloc can take upwards of 2ms, but these disappear as the program settles into a steady state.

\begin{figure}
    \centering
    \includegraphics[width=\linewidth]{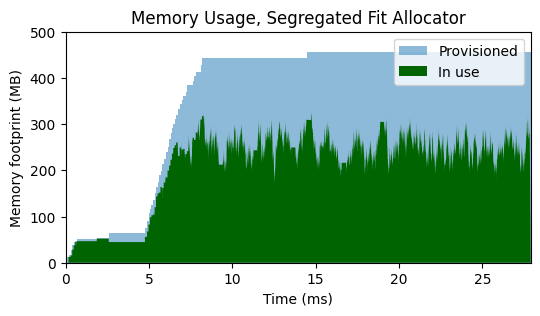}
    \caption{Memory usage during benchmark}
    \label{fig:mem_usage}
\end{figure}

We measure fragmentation in our experiments (Figure \ref{fig:mem_usage}) by tracking not only the memory in use by the benchmark, but also the physical device memory provisioned to the benchmark process. In our segregated fit allocator, we observe a roughly 50\% fragmentation ratio. This is consistent with prior work\cite{johnstone1998memory} on segregated fit. A Two-Level Segregated Fit allocator would see improved fragmentation\cite{masmano2004tlsf}.

\section{Conclusion}\label{sec:conclusion}

We note that prior work on dynamic memory allocation relies on free lists and boundary tags. We've posed a heterogeneous system model where memory is stored on one device and managed from another. After observing that existing algorithms are no longer applicable, we've provided alternatives to free lists and updated versions of classic sequential and segregated fit algorithms.

An implementation of our segregated fit algorithm was tested against conventional calls to cudaMalloc and cudaFree. It was found that, after a warmup period, our work consistently outperformed the CUDA calls by two orders of magnitude. This is mainly due to its implicit and efficient memory recycling. 

We do not claim to outperform established GPU allocators \cite{winter2021dynamic} at scale, nor to address their highly parallelized use-cases. However, in specific circumstances where kernels are not expected to allocate memory and diverse accelerated computation is combined into a large heterogeneous application, performance benefits can be achieved by managing memory on the host. It is unnecessary to invoke device calls to free memory, and instead it can be dynamically recycled.

\section*{Acknowledgements}
This work was sponsored by NSF Grant 2229290

This work was edited with the assistance of ChatGPT

\bibliographystyle{IEEEtran}
\bibliography{references}

\clearpage
\appendix

\section{Hybrid-Array Lists}
\label{app:hal}

A hybrid array list (HAL) is  a series of fragmented chunks of an array linked together in a linked list. This allows for accelerated lookup times while retaining the O(1) insertion and deletion times of conventional linked lists. Incorporating hashtables can reduce lookup time to O(1) as well.

\begin{figure}
        \centering
        \includegraphics[width=\linewidth]{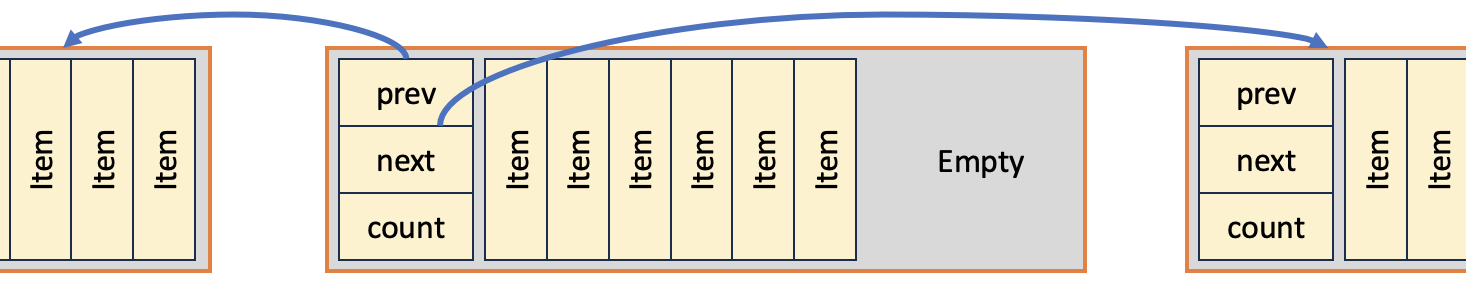}
        \caption{Structure of Hybrid Array List}
        \label{fig:hal_design}
\end{figure}

Each chunk is not assumed to be full, but it is assumed to have contiguous entries. The entries are references to free blocks and are stacked in the front half of the chunk, with the rear half leaving room for expansion. The entries are assumed to be sorted. A diagram of our HAL design is presented in Figure~\ref{fig:hal_design}.

Insertion and removal operations do not trigger a full sort within the chunk. Since the chunk is already sorted, all entries after that point are shifted to create or occupy the empty space. The insertion/deletion point can be easily identified with a O(logn) binary search of the chunk.

If insertion would cause the chunk to overflow, then it is split in two. If a removal would cause the chunk to be empty, then it is removed from the larger linked list.

As in prior work, the time complexity of insertion and removal operations from Hybrid Array Lists scales only on the size of the array chunk, not the length of the entire list.

An algorithm is shown for initialization, Algorithm \ref{alg:hal_init}. Interactions with the HAL are described below.

\subsection{Insert}

Once an insertion point has been identified, every item in the chunk at and past that point must be shifted over. If the chunk is already completely full, it is first split, with half its contents going to a new chunk.

After the shifting process, the item is inserted into its selected location. Insertion time is linear with respect to the maximum size of the chunk, but does not scale with the size of the wider list.

\subsection{Remove}

To remove an item from a chunk, every item after that point is shifted over to overwrite it.

If the item to be removed is the last item in the chunk, the chunk is removed from the HAL by standard linked-list procedure.

\subsection{Search}
It is worth noting that, for our memory management purposes, search is done in reverse order. Given that $x$ is the requested deallocation address, $y_d$ is the corresponding block to be deallocated, and $y$ is the set of all blocks, we say
\\
$$y_d = \max(y_i \in y) \mid y_i \leq x$$
\\
As we iterate over the free list, we check the first (smallest) entry in each chunk. If it is larger than $x$, clearly $y_d$ is not present, and we can continue checking the next previous chunk. Once an entry $y_i \leq x$, then we know that $y_d$ is somewhere in the current chunk. At this point, a simple binary search can identify it.

In a more general sense, search over a sorted HAL is done linearly over each chunk until one is confirmed to contain the element. Then a binary search is done within the chunk.

\subsection{Iterators}

As with any container, we can define iterators for a hybrid-array list to abstract the container into an ordered range. An iterator is implemented as the pair of a chunk reference and an index within that chunk.

The increment operation is normally as simple as incrementing the index. However, if doing so would push it past the end ($index \geq count$), then we perform

\begin{equation}
\begin{gathered}
chunk \gets chunk.next \\
index \gets 0
\end{gathered}
\end{equation}

Decrement operations are very similar but in reverse. The index is normally decremented, but if doing so would cause it to be negative, then we perform

\begin{equation}
\begin{gathered}
chunk \gets chunk.prev \\
index \gets chunk.count - 1
\end{gathered}
\end{equation}

\begin{algorithm}
\caption{Initialize Hybrid-Array List}\label{alg:hal_init}
$free\_list\_head \gets \text{alloc page}$
$used\_list\_head \gets \text{alloc page}$
$free\_list\_head.count \gets 1$
$free\_list\_head.prev \gets NULL$
$free\_list\_head.next \gets NULL$
$free\_list\_head.blocks[0].address \gets \text{heap start}$
$free\_list\_head.blocks[0].size \gets \text{heap size}$
$used\_list\_head.count \gets 0$
$used\_list\_head.prev \gets NULL$
$used\_list\_head.next \gets NULL$
\end{algorithm}

\end{document}